\documentclass[aps,prl,reprint,superscriptaddress,10pt,nobibnotes]{revtex4-2}

\usepackage{graphicx,amsmath,amssymb,dcolumn,physics,color,wasysym,lipsum}

\pdfoutput=1

\usepackage{graphicx}%
\usepackage{dcolumn}%
\usepackage{bm}%
\usepackage[colorlinks= true, linkcolor=blue, citecolor=blue, urlcolor=blue]{hyperref}%
\usepackage{lipsum}
\usepackage{mathtools}
\usepackage{xcolor}
\usepackage{multirow}
\usepackage{graphicx,amsmath,amssymb,mathtools, dcolumn,physics,color,wasysym, bm}
\usepackage{booktabs}

\graphicspath{{./figs/}{./}}

\begin{document}

\title{A structure-based model of the bacterial chemosensory array reveals how the scaffolding protein CheW tunes sensory responses}

\author{David Hathcock}
\email{david.hathcock@utoronto.ca}
\affiliation{Department of Physics, University of Toronto, Toronto, Ontario M5S 1A7, Canada}

\author{Yingxin Zheng}
\affiliation{Lewis-Sigler Institute for Integrative Genomics, Princeton University, Princeton, NJ 08544, USA}

\author{Yuhai Tu}
\email{ytu@flatironinstitute.org}
\affiliation{Center for Computational Biology \& Center for Computational Neuroscience, Flatiron Institute, New York, NY 10010, USA}

\begin{abstract}

We introduce a structure-based Ising model of the bacterial chemosensory array to investigate how the scaffolding protein CheW tunes signaling responses. The model couples the activity states of core signaling units and of CheW subunits within hexameric rings on a lattice with the structure observed in cryoEM experiments. Recent experiments show that increasing CheW participation in hexameric rings raises response cooperativity in non-adapting cells and sensitivity in adaptation-proficient cells. Our lattice model quantitatively reproduces both effects, while mean-field theory shows that CheW increases the effective coupling between core units, sharpening the response in both cell types. The statistics of CheW placement (complete hexameric rings versus random distribution) substantially alter how responses vary with CheW abundance. Moreover, different coupling strengths can yield the same static response yet markedly different activity fluctuations and dynamics. CheW abundance further controls a crossover from relaxational to oscillatory fluctuations, whose location depends on the spatial distribution of CheW. These dynamical signatures, including oscillations predicted near native CheW levels, are testable by single-cell dynamics measurements. Combined with CheW titration and our modeling results, these measurements can probe the array's assembly dynamics and allosteric couplings. Beyond the chemosensory array, our approach can generalize to other large protein complexes, providing a framework to study the functional role of scaffolding proteins and allosteric interfaces in signal processing.

\end{abstract}

\maketitle

\section{Introduction}

Protein complexes in biology often exhibit cooperative responses: binding of substrates at multiple sites leads to sharp changes in protein conformation and the resulting downstream biological signaling and function~\cite{monod1965natureallosteric, berg2003rotarymotor, edelstein1971extensionsallosteric, changeux2005allostericmechanisms}. Cooperativity is particularly apparent in the sensory pathway for bacterial chemotaxis: the chemosensory array in \emph{Escherichia coli} is composed of thousands of transmembrane receptors, histidine kinase enzymes, and coupling proteins that form an extended lattice structure~\cite{parkinson2015signalingsensory}. It has long been appreciated that coupling between signaling units is the source of cooperativity in the chemosensory array: both the spatially extended Ising-like models~\cite{bray1998receptorclustering, duke1999heightenedsensitivity, mello2003quantitativemodeling, mello2004effectsreceptor, lan2011adaptlocally} and the simplified  all-or-none Monod-Wyman-Changeux (MWC) model~\cite{sourjik2004functionalinteractions,mello2005allostericmodel,keymer2006chemosensingescherichia,shimizu2010modulargradientsensing} have provided successful quantitative theories for capturing chemosensory responses. More recent work has generalized these models of cooperativity to address the role of nonequilibrium dynamics (i.e. phosphorylation-dephosphorylation reaction cycles) in signal propagation. Specifically, nonequilibrium allostery~\cite{hathcock2023nonequilibriumallosteric} and nonequilibrium lattice~\cite{hathcock2024timereversalsymmetry,sherry2026latticeultrasensitivity} models have been respectively introduced to explain measured discrepancies between binding and kinase response~\cite{amin2010chemoreceptorssignalling, levit2002receptormethylation} and irreversibility observed in the fluctuations of kinase activity~\cite{keegstra2026spontaneousswitching}.

\begin{figure*}[t]
   \includegraphics[width=\textwidth]{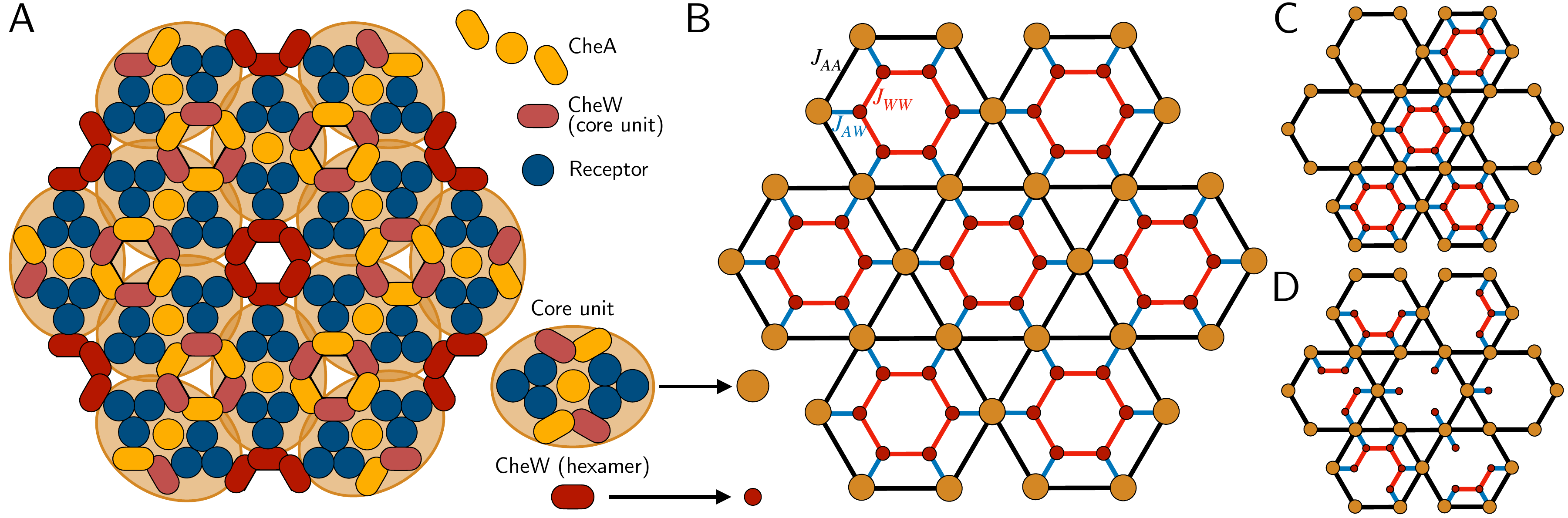}
   \caption{\emph{A structure-based model of the chemosensory array.} (A) Schematic of the chemosensory array structure, showing the core unit (two trimers of receptor dimers -- blue, two CheW -- light red, and one CheA -- yellow) and the hexameric ring of CheW (red) that fills the holes in the Kagome lattice. Shaded ovals outline individual core units. (B) Coarse-grained spin-lattice model of the array, where each site corresponds to a core unit (dark yellow) or a CheW in a hexamer (red). Spins interact allosterically with coupling strengths $J_{AA}$, $J_{WW}$, and $J_{AW}$. (C-D) Two scenarios for how CheW are distributed in the array when CheW expression is intermediate $0<f<1$: (C) CheW preferentially form hexamers, filling a fraction $f$ of the holes in the Kagome lattice; (D) CheW are randomly distributed across all possible sites in the array with filling fraction $f$.}
   \label{fig:structure_model}
\end{figure*}

Despite this progress, there has been limited modeling work~\cite{khursigara2011lateraldensity} studying the role played by the microscopic protein structure of the chemosensory array in its functional response. High-resolution cryogenic electron microscopic (cryoEM) imaging has revealed the detailed structure of the array~\cite{briegel2009universalarchitecture,liu2012moleculararchitecture,briegel2012bacterialchemoreceptor,cassidy2015cryoemcomputer, cassidy2020structuredynamics}: two trimers of receptor dimers, two CheW scaffolding proteins, and one histidine kinase CheA dimer form a core unit. Core units couple together via the CheA-CheW interface-2 interaction to form an extended array with Kagome lattice structure, and additional CheW can form hexamers that fill the holes in this lattice. The result (Fig.~\ref{fig:structure_model}A) is a highly structured array with multiple distinct interfaces of interaction between its protein components, yet the quantitative role this structure plays in shaping the array's functional response remains unclear.

One main limiting factor is that CheW's conformational states and allosteric couplings are especially difficult to pin down experimentally. Unlike for the receptors and the kinase CheA, the activity of the internal scaffold protein CheW cannot be probed directly. Structure-based modeling is therefore essential for inferring CheW's functional role, since its responses and interactions can only be accessed indirectly through its effects on the array's collective signaling response.

Recent experiments have examined the role played by CheW hexamers in chemotaxis signaling by varying CheW expression levels and measuring the resulting chemosensory responses~\cite{pinas2022hexamericrings} in both adaptation-proficient (\emph{cheRB})$^+$ and non-adapting $\Delta$(\emph{cheRB}) mutant cells. These measurements revealed a striking contrast between the two cell types. In non-adapting cells, increasing CheW ring content increases the cooperativity (Hill coefficient) of the response but does not considerably change the absolute sensitivity ($K_{1/2}$, half-maximum concentration). In adaptation-proficient cells, increasing CheW ring content has the opposite effect: cooperativity is roughly maintained while sensitivity increases considerably ($K_{1/2}$ decreases).

Motivated by these intriguing results, in this paper we introduce a structure-based Ising model of the chemosensory array. We model both the activity states of core units and CheW proteins participating in the hexameric rings, with a biologically accurate protein interaction structure. Our model incorporates three allosteric coupling interactions: between neighboring core units, between neighboring CheW in a hexamer, and between CheW and the core unit. It also includes local fields that capture the effects of ligand binding and methylation on core unit activity. By studying the model via lattice simulations and mean-field theory, we identify the microscopic mechanisms underlying the experimentally measured effect of CheW on the signaling response of non-adapting and adapting cells. Increases in the effective coupling between core units due to CheW hexamers sharpen the response curve in both cell types; for non-adapting mutants that are active at zero ligand concentration, the sharper response leads to an increase in cooperativity, while for adaptation-proficient cells that have adapted to intermediate activity, the sharper response leads to an increase in sensitivity.

Finally, our model allows us to investigate how the statistics of CheW placement and strengths of couplings between array components quantitatively affect chemosensory responses and dynamics. Comparing two array assembly scenarios, in which (1) CheW preferentially form hexamers and (2) CheW are randomly distributed in open binding sites, we find considerable differences in how response properties depend on CheW expression. Activity dynamics and temporal correlations also depend strongly on the internal allosteric couplings within the array, even among sets of couplings that have identical static, time-averaged responses. CheW abundance further influences dynamics, inducing a transition from  relaxational fluctuations with one or two timescales to adaptation-induced oscillations.

Overall, our results highlight the importance of the structural details of the chemosensory array in determining its functional responses, motivate future experiments measuring chemosensory responses and single-cell dynamics across different CheW expression levels, and suggest that CheW may play a role in tuning the sensitivity of the array via long-timescale structural adaptation. Beyond the chemosensory array, our structure-based modeling approach provides a general framework for studying the function of scaffolding components like CheW, whose internal conformational states are difficult to probe experimentally but  contribute significantly to the signaling responses of a large protein complex.

\section{Results}

\subsection{A structure-based model of the chemosensory array}

To model the response of the chemosensory array with varying CheW expression we consider a coarse-grained two-species spin model (Fig.~\ref{fig:structure_model}). Spins $a_i$ represent the CheA kinase activity of the core unit and $w_i$ represent the CheW conformation ($a_i$,$w_i=0$: inactive or 1: active). The $a$-$a$ interaction $J_{AA}$ acts along the interface-2 connections (which form a Kagome lattice). The $w$-$w$ interaction $J_{WW}$ acts between CheW in the hexameric ring. Finally, the $w$-$a$ interaction $J_{AW}$ acts between each hexamer CheW and the closest CheA on the Kagome lattice. There are two fields that act on the core unit spins: $h_L$ is a global field determined by binding of extracellular ligands to the receptors and $h_M$ is a local field determined by the average methylation level of the receptors in the core unit. Together these interactions give us the following Hamiltonian
\begin{equation}
   \begin{split}
      H &= -J_{AA} \sum_{\langle i, j \rangle} s_i s_j - J_{WW}\sum_{\langle i, j \rangle} \sigma_i \sigma_j \\
      &- J_{AW}\sum_{\langle i, j \rangle} \sigma_i s_j
      + \frac{1}{2}\sum_i \big [ h_L([L]) + h_M(m_i) \big]s_i,\label{eq:model_hamiltonian}
   \end{split}
\end{equation}
where $s_i= 2 a_i-1$ and $\sigma_i= 2 w_i-1$ are Ising spins ($s_i,\sigma_i=\pm 1$), and the sums are over neighbors in the structured array.  We use an MWC-like ligand field $h_L([L]) = \log [(1+[L]/K_i)/(1+[L]/K_a)]$, where $K_i$ and $K_a$ are the ligand dissociation constants for the inactive and active states of the core unit, respectively. This choice is consistent with experiments showing that isolated core unit responses have Hill coefficients close to 1~\cite{li2014selectiveallosteric,pinas2016sourcehigh}. In non-adapting mutants the methylation field is fixed to $ h_{QEQE}$. Following previous studies~\cite{shimizu2010modulargradientsensing, lan2011adaptlocally}, the methylation field in adapting cells is modeled as a linear function of the local methylation level $h_M(m_i) = -\alpha (m_i - m_0)$, where $m_i$ is the average methylation level of the receptors in the core unit, $\alpha$ is the free energy change per added methyl group, biasing the core unit toward the active state, and $m_0$ is the methylation level at which the core unit is unbiased between active and inactive states (at zero ligand concentration).

Because the experiments of interest vary CheW expression level, a fraction $f$ of the total possible CheW sites in Fig.~\ref{fig:structure_model}B are filled. When $f<1$, the sums in the Hamiltonian, Eq.~(\ref{eq:model_hamiltonian}), only run over filled CheW sites. For simplicity, we consider two extreme scenarios for how CheW are placed in the array: CheW either preferentially form hexameric rings, which fill a fraction $f$ of the holes in the Kagome lattice [Fig.~\ref{fig:structure_model}C], or CheW are randomly distributed across all possible sites in the array [Fig.~\ref{fig:structure_model}D].

Many parameters in the model can be estimated from experiments and previous modeling work; specifically $K_i, K_a, h_{QEQE}, \alpha, m_0$ are fixed throughout our study (see Table~\ref{tab:parameters}). We focus on how the response changes with CheW expression level $0<f<1$ and the distribution of local methylation levels $P(m_i)$, which differs between non-adapting and adapting cells. The coupling parameters $J_{AA}, J_{WW}, J_{AW}$, which describe the strength of allosteric interactions between different protein components of the array, are not well characterized experimentally; we explored the space of coupling parameters in numerical simulations to identify regimes which quantitatively reproduce the experimentally measured response behaviors. The primary set of couplings used throughout our results is given in Table~\ref{tab:parameters}.

\begin{figure*}[t]
   \includegraphics[width=0.75\textwidth]{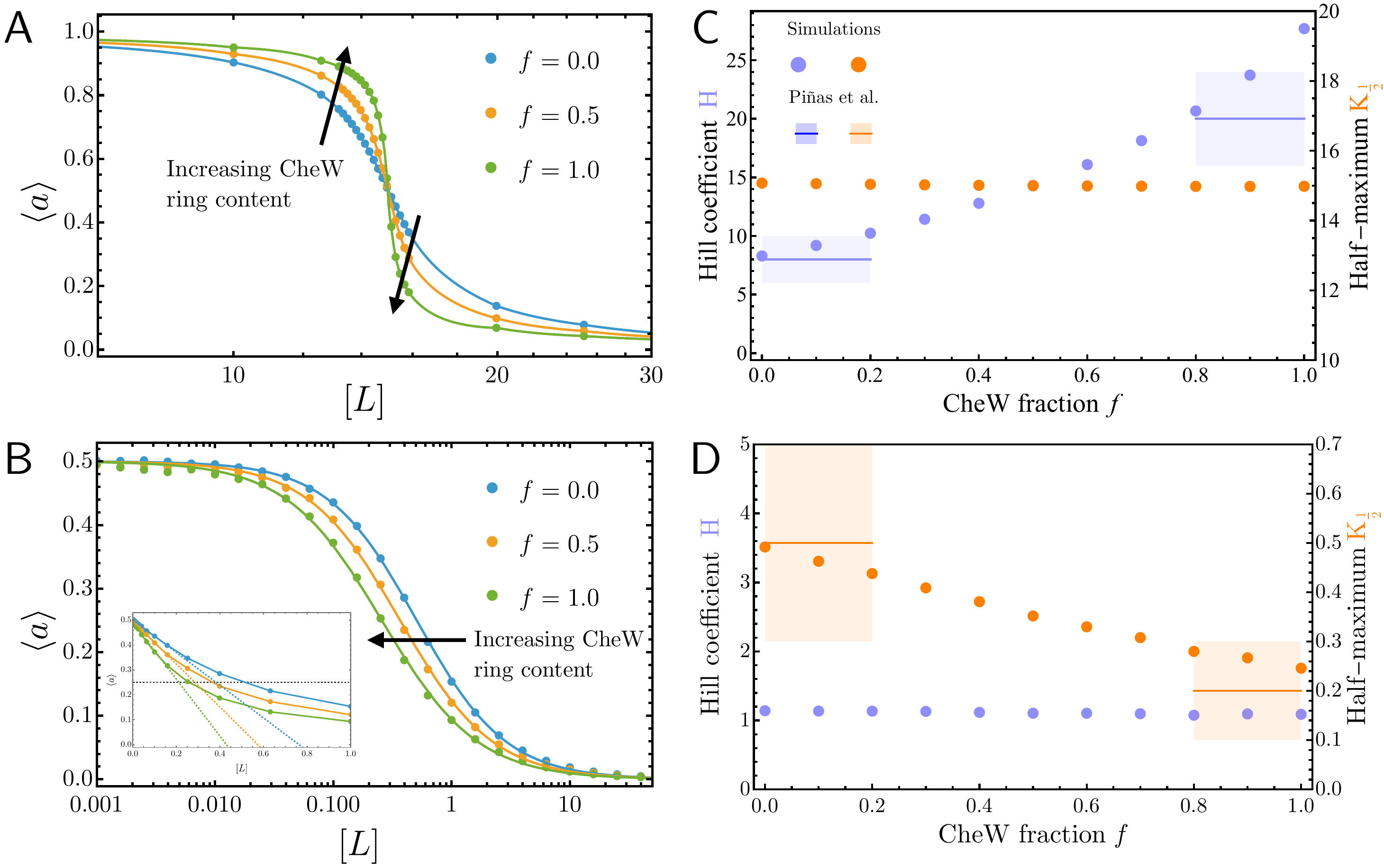}
   \caption{\emph{Dependence of signaling response properties on CheW fraction $f$.} 
   (A-B) Dose-response of the non-adapting QEQE model (A) and the adaptation model (B) for different CheW fractions. Adding more CheW to the array increases the cooperativity of the response in QEQE cells and increases the sensitivity in adaptation-proficient cells. The inset in (B) shows a linear fit (dashed lines) to the response at zero concentration, analogous to our mean-field result, Eq.~(\ref{eq:K12}). Solid lines are interpolations of the simulation data in (A) and Hill function fits in (B).
   (C-D) The Hill coefficient $H$ and half-maximum concentration $K_{1/2}$ as a function of CheW fraction $f$ for the non-adapting (C) and  adapting (D) models. Shaded regions show the Pi\~nas et al.~\cite{pinas2022hexamericrings} measurements; since the CheW fraction is not precisely known in experiments we show ranges $[0,0.2]$ and $[0.8,1]$ for the low and native CheW expression measurements respectively. 
   }\label{fig:responseProperties}
\end{figure*}

We simulate the model using a Metropolis Monte Carlo algorithm and measure the activity response as a function of ligand concentration $[L]$ for different CheW fractions $f$ and methylation distributions $P(m_i)$. Coupling $J_{AA}$ is fixed by matching the response properties in $f=0$ simulations to the low CheW expression experiments, while $J_{AW}$ and $J_{WW}$ are chosen to fit native CheW expression measurements~\cite{pinas2022hexamericrings} (see Methods for details of simulation and parameter selection). As detailed further below $(J_{AW}, J_{WW})$ are not uniquely constrained by experiments. We also developed a mean-field theory for the model, which allows us to derive the analytical expressions given below for the Hill coefficient and $K_{1/2}$ (see Methods). These expressions provide an intuitive picture for how CheW combines with the methylation distribution to enhance the cooperativity and sensitivity of the array in different cell types.

\begin{table}[tbp]
    \centering
    \caption{\emph{Lattice Model Parameters.} Listed couplings $(J_{AW},J_{WW})$ are used throughout all results except Fig.~\ref{fig:fluctuations}, where multiple couplings are considered, and Fig.~\ref{fig:fluctuationsCheWlevel} where we use $(J_{AW},J_{WW})=(0.35, 0.375)$. }
    \label{tab:parameters}
    \begin{tabular}{llcl}
        \toprule
        \textbf{Parameter} &  & \quad\quad \textbf{Value} & \\
        \midrule
        Inactive dissociation constant  &\quad $K_i$            &   \quad \quad 2 $\mu$M                \\
        Active dissociation constant  &\quad $K_a$            &    \quad \quad 10,000 $\mu$M            \\
        Mutant methylation field  &\quad $h_{QEQE}$            &     \quad \quad  $-2.14  k_B T$             \\
        Methylation energy scale  &\quad $\alpha$            &     \quad \quad  1 $ k_B T$             \\
        Crossover methylation level  &\quad $m_0$            &                  \quad \quad 0.75  \\
        Core unit-core unit coupling  &\quad $ J_{AA}$            &     \quad \quad  0.35 $k_B T$             \\
        Core unit-CheW coupling  &\quad $ J_{AW}$            &     \quad \quad  0.25 $k_B T$             \\
        CheW-CheW coupling  & \quad $ J_{WW}$            &     \quad \quad  0.565 $ k_B T$             \\
        \bottomrule
    \end{tabular}
\end{table}

\subsection{CheW-mediated coupling in non-adapting mutants strengthens effective array cooperativity}

For the non-adapting mutant cells, all receptors are in the QEQE methylation state, and the methylation level is fixed and homogeneous across the array, $P(m_i) = \delta_{m_i,2}$. In this case the activity is $\langle a \rangle([L]=0)\approx 1$ at zero ligand concentration, and the activity is fully suppressed at saturating ligand, $\langle a \rangle([L]\to\infty) \approx 0$. Thus $K_{1/2}$ corresponds to $\langle a \rangle = 1/2$ (equivalently $\langle s \rangle = 0$), which occurs when the total field $h_\mathrm{tot} = h_L + h_M=0$. The zero-field condition is independent of the amount of CheW in the array, so $K_{1/2}$ is independent of CheW fraction $f$ as seen in the Pi\~{n}as et al. experiments~\cite{pinas2022hexamericrings}. On the other hand, we expect the Hill coefficient to increase with CheW fraction $f$ because the addition of CheW hexamers in the array provides additional coupling pathways between core units.

Our lattice simulations confirm these qualitative behaviors seen in experiments: the Hill coefficient increases considerably with CheW fraction $f$ while $K_{1/2}$ remains roughly constant. Fig.~\ref{fig:responseProperties}A shows the simulated response curves for different CheW fractions $f$, while Fig.~\ref{fig:responseProperties}C shows the Hill coefficient and $K_{1/2}$ extracted from the response curves. By tuning the coupling parameters $J_{AA}, J_{AW}, J_{WW}$, we can quantitatively match the experimentally measured response curves and Hill coefficients from Pi\~{n}as et al.~\cite{pinas2022hexamericrings}.

Our mean-field theory provides further insight into the mechanisms by which CheW hexamers enhance the cooperativity of the array. We perturbatively compute the response near $K_{1/2}$ (or $h_\mathrm{tot}=\langle s\rangle=\langle \sigma \rangle=0$) to derive an expression for the Hill coefficient $H = -2\eval{\partial \log \langle a \rangle / \partial \log [L]}_{[L]=K_{1/2}}$ (Methods). We find the Hill coefficient is given by,
\begin{equation}\label{eq:Hill_nonadapt}
    H = \chi_h \cdot \frac{1}{1-4 J_{AA} - \frac{2 f J_{AW}^2}{1-2 J_{WW}}},
\end{equation}
where the first factor $\chi_h = \eval{\partial h_L/\partial \log[L]}_{[L]=K_{1/2}}$ is the sensitivity of the ligand field $h_L$ to ligand concentration at $K_{1/2}$, which is always $\leq 1$. From this expression, we can identify the effective coupling $J_{AA}^\text{eff} = J_{AA} + \frac{1}{2}\frac{f J_{AW}^2}{1-2 J_{WW}}$ between core units, which increases with both CheW fraction $f$ and the strengths of the allosteric couplings, $J_{AW}$ and $J_{WW}$. The increase in coupling, $\frac{1}{2}\frac{f J_{AW}^2}{1-2 J_{WW}}$, captures the new coupling pathways between neighboring core units (e.g. $A$-$W$-$W$-$A$), which occur with frequency $f$ and involve two $A$-$W$ interactions ($J_{AW}^2$) coupling the core units to the CheW ring. The $(1-2 J_{WW})^{-1}$ term captures the overall coupling strength introduced by the cooperativity of the CheW subunits in the hexameric ring. This calculation thus provides a quantitative picture for how CheW rings enhance the cooperativity of the array. Here we have assumed CheW preferentially form hexamers, but a similar calculation applies to randomly distributed CheW, leading to the same result with the substitution $J_{WW} \to f J_{WW}$ so that $J_{AA}^\text{eff} = J_{AA} + \frac{1}{2}\frac{f J_{AW}^2}{1-2 f J_{WW}}$. The additional factor $f$ accounts for the fact that $W$-$W$ interactions only occur with frequency $f$ when CheW are randomly distributed. The differences in response between the two CheW binding scenarios will be discussed further below.

\subsection{CheW-mediated coupling in adapting cells enhances sensitivity }

For adaptation-proficient cells, the methylation level of each core unit is controlled by a feedback mechanism facilitated by methylation and demethylation reactions, allowing the cells to adapt to an average activity $\langle a \rangle([L]=0)=A_0\approx 1/3$--$1/2$~\cite{tu2008modelingchemotactic}. Since the adaptation timescales are slow compared to the response timescales, we assume the methylation level of each core unit is fixed during the response measurement, but varies across the array according to a distribution $P(m_i)$, leading to a quenched random field $h_M(m_i)$ acting on each core unit.

\begin{figure*}[t]
   \includegraphics[width=\textwidth]{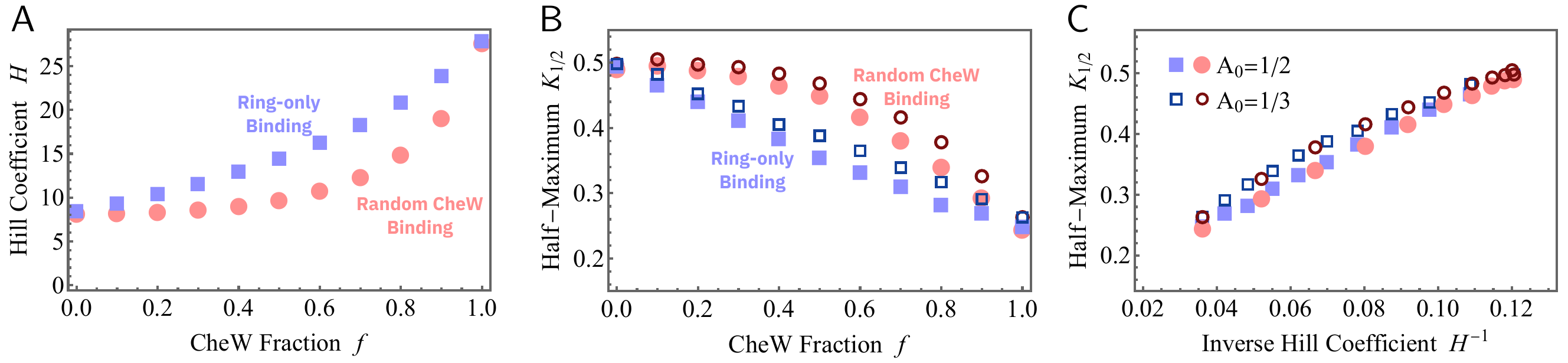}
   \caption{\emph{Response properties for intermediate CheW fraction depend strongly on ring formation statistics.} (A-B) The Hill coefficient $H$ in the non-adapting model (A) and the half-maximum concentration $K_{1/2}$ in the adapting lattice model (B), each plotted as a function of CheW fraction $f$ for two ring formation scenarios: ring-only binding, where all CheW are placed in full rings in the array, and random binding, where CheW placement has no spatial correlation. (C) Plotting $K_{1/2}$ versus $H^{-1}$, we see a linear collapse of the data as predicted by mean-field theory. In (B-C) we show results for adaptation to both $A_0=1/2$ (solid symbols) and $A_0=1/3$ (open symbols).
   }\label{fig:RingvsRandom}
\end{figure*}

We determine the adapted methylation distribution $P(m_i)$ in our lattice simulations by simulating the stochastic methylation and demethylation dynamics based on the local activity state of each core unit (see Methods and Ref.~\cite{mello2004effectsreceptor}). This simulation is used to measure the steady-state methylation distribution $P(m_i)$, which is then fixed as a quenched random field  across core units in our simulations of response functions. Fig.~\ref{fig:responseProperties}B shows a few response curves for the adaptation model and Fig.~\ref{fig:responseProperties}D shows the Hill coefficient and $K_{1/2}$ extracted from response curves for several values of $f$. In quantitative agreement with experiments~\cite{pinas2022hexamericrings}, we find that increasing CheW fraction $f$ increases the sensitivity of the response (decreases $K_{1/2}$) while maintaining a roughly constant Hill coefficient. This qualitative result is robust to changes in the methylation distribution, e.g.  the distribution $P(0)=P(1)=1/2$ differs from the simulated methylation distribution, but also yields an adapted activity of $A_0 \approx 1/2$ and similar changes in $K_{1/2}$ with $f$. This result shows that adaptation to an intermediate activity level, rather than details of the methylation distribution, is the key feature underlying measured changes in $K_{1/2}$. 
From an adapted state, increases in coupling via CheW participation generically lead to enhanced ligand sensitivity.

Our mean-field model provides further insight into the mechanisms for sensitivity enhancement. We can estimate the change in $K_{1/2}$ with CheW fraction $f$ by computing the linear perturbation to the response near $[L]=0$. Assuming the adapted activity at zero ligand concentration is $A_0$, we find the following expression for $K_{1/2}$,
\begin{equation}\label{eq:K12}
   K_{1/2} \approx 2 K_i A_0 \left [1/\chi_s-4 J_{AA}-\frac{2 f \chi_\sigma J_{AW}^2}{1-2 \chi_\sigma J_{WW}} \right],
\end{equation}
where $K_i$ is the dissociation constant for the inactive state, $\chi_s = \sum_m P(m) \sech^2 [h_{MF,s_0}(m)]$ and $\chi_\sigma = \sech^2 (h_{MF,\sigma_0})$ are the susceptibilities at adapted activity $A_0$ for the core unit and CheW spins, respectively, and $h_{MF,s_0}(m)$ and $h_{MF,\sigma_0}$ are the mean fields acting on the core units (with methylation $m$) and CheW spins at zero ligand concentration (see Methods for full expressions). %

Since the adapted state has tuned the array toward intermediate activity, near the sensitive regime of the response curve, increasing coupling via CheW sharpens the response at the background concentration (in this case $[L]=0$). The result is that the half-maximum concentration $K_{1/2}$ decreases, corresponding to the observed increase in response sensitivity. Equation~(\ref{eq:K12}), demonstrates how coupling decreases $K_{1/2}$ and again allows us to identify an effective coupling between core units, $J_{AA}^\text{eff} = J_{AA} + \frac{1}{2}\frac{f \chi_\sigma J_{AW}^2}{1-2 \chi_\sigma J_{WW}}$. Similar to the non-adapting cells, the effective coupling increases with CheW fraction due to the new coupling pathways between core units. In this case the couplings involving CheW are weighted by the susceptibility $\chi_\sigma \leq1$ of the CheW spins in the adapted state. As in the non-adapting model our results hold analogously for randomly distributed CheW with the substitution $J_{WW} \to f J_{WW}$, so that $J_{AA}^\text{eff} = J_{AA} + \frac{1}{2}\frac{f \chi_\sigma J_{AW}^2}{1-2 \chi_\sigma f J_{WW}}$.

Though the invariance of the Hill coefficient is difficult to show analytically even within the mean-field model, the classical MWC model provides intuition for this result. Specifically, with average methylation level tuned so that the activity is $A_0$ at zero ligand concentration, the MWC model predicts that the Hill coefficient at the half-maximum, $A_0/2$, is approximately $H\approx (2-A_0)\log[(2-A_0)/(1-A_0)]$ as long as the correlation length $N$ or the binding-conformation coupling $\log(K_a/K_i)$ is sufficiently large (see Methods). Thus, if the effective coupling in our lattice model is strong enough to maintain a large correlation length, we expect the Hill coefficient to be approximately invariant to changes in CheW fraction $f$, as found in our simulations.

\subsection{Statistics of CheW ring formation modulate response properties}

Beyond explaining available experimental data, our model provides predictions for how different CheW hexamer assembly dynamics affect the array's response properties. We have considered two extreme binding scenarios: (1) CheW preferentially form hexameric rings, which fill a fraction $f$ of the holes in the Kagome lattice, or (2) CheW are randomly distributed across a fraction $f$ of all possible open binding sites in the array. For the limits $f=0$ and $f=1$, the two scenarios are identical: the array has no CheW in hexameric binding sites or all CheW binding sites are filled, respectively. These limits roughly correspond to the regimes probed by established experiments~\cite{pinas2022hexamericrings}: response was measured at CheW expression at half-native and native levels. For the former, almost all CheW are used to form core units and hexameric rings are largely absent, occurring at $\sim 5\%$ of the rate seen in native arrays. At native levels, CheW is roughly two-fold in excess of the core-unit stoichiometry~\cite{li2004cellularstoichiometry, pinas2022hexamericrings}, suggesting there may be sufficient CheW to fill most hexameric binding sites.

For intermediate values of $f$, however, our simulations reveal that the two binding scenarios lead to markedly different response properties. Fig.~\ref{fig:RingvsRandom} shows the dependence of the Hill coefficient (A) and $K_{1/2}$ (B) on CheW fraction $f$ for ring-preferring and randomly bound CheW. We find that the change in response properties for ring-preferring CheW is more consistent across the full range of $f$; indeed $H^{-1}$ and $K_{1/2}$ are both approximately linear in $f$ (except for very small $f$), as predicted by our mean-field theory, Eqs.~(\ref{eq:Hill_nonadapt}) and (\ref{eq:K12}). In contrast, random CheW binding leads to a gradual, nearly quadratic change in response properties for small values of $f$, with a more dramatic change as $f$ approaches 1. This difference arises because at small $f$, additional CheW are isolated so new coupling pathways between core units are rare to leading order in $f$. The increase in effective coupling strength due to these pathways is weak, $J_{AW}^2 \ll 1$ and $1-2f J_{WW} \approx 1$, whereas for ring-preferring assembly the denominator $1-2 J_{WW}$ is smaller than 1 due to the cooperativity of the CheW in the hexameric ring.

Independent of array assembly dynamics, our mean-field theory predicts a collapse between the Hill coefficient in non-adapting cells and $K_{1/2}$ in adaptation-proficient cells to an approximately linear relationship, $K_{1/2} \approx c H^{-1} + d$, where $c$ and $d$ are constants that depend on the coupling parameters, the methylation distribution, and adapted activity $A_0$ (see Methods). This collapse is exact if the susceptibility $\chi_\sigma=1$, which occurs when the adapted activities are $s_0=\sigma_0=0$ at zero ligand concentration. The collapse demonstrates that the same underlying mechanism, i.e. CheW-induced coupling pathways between core units, is responsible for the changes in both the Hill coefficient and $K_{1/2}$ in non-adapting and adapting cells, respectively. Our lattice simulations confirm that this collapse also occurs in two-dimensional arrays (Fig.~\ref{fig:RingvsRandom}C), holding nearly exactly for adaptation to both $A_0=1/2$ and $1/3$. Interestingly, contrary with the mean-field prediction ($K_{1/2} \propto A_0$), decreasing $A_0$ increases $K_{1/2}$ for a given $f$. We conjecture this discrepancy is due to spatial correlations in methylation between neighboring core units in the lattice model. For $A_0=1/3$ we observe weaker anti-correlations than for $A_0=1/2$; anti-correlations appear to play a significant role in setting the response sensitivity.

\begin{figure*}
   \includegraphics[width=0.78\textwidth]{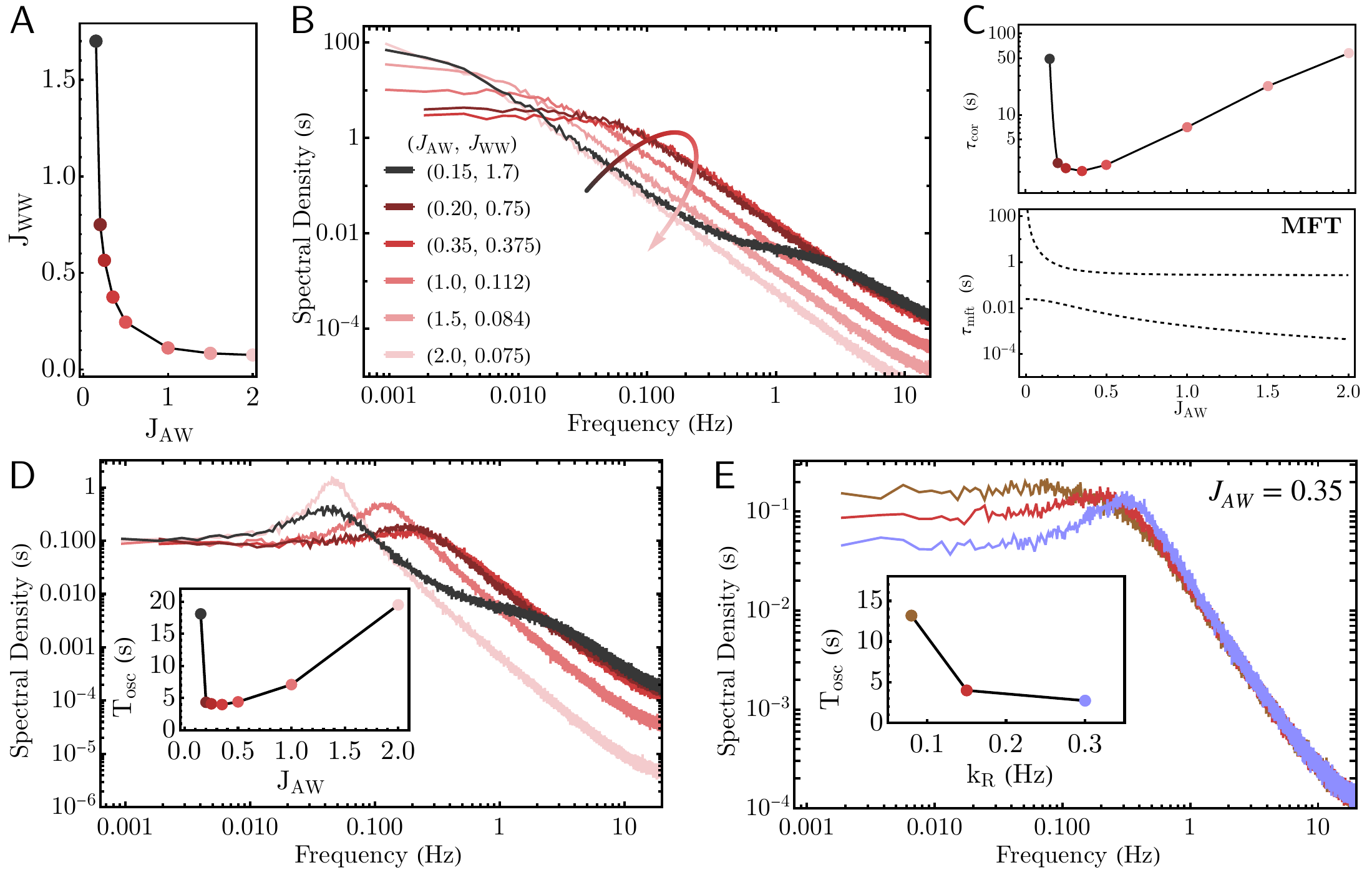}\\
   \caption{\emph{Relative strengths of allosteric couplings modulate activity dynamics of the chemosensory array.} 
   (A) The curve $(J_{AW}, J_{WW})$ in coupling space that produces the same response Hill coefficient $H=27$. 
   (B) The power spectral density of activity fluctuations in the non-adapting model for different coupling pairs with the same static response.
   (C) Upper: The dominant correlation time in (B), which is non-monotonic as we vary $(J_{AW}, J_{WW})$ along the curve of constant Hill coefficient. 
   Lower: Our mean-field theory predicts the initial decrease in correlation time with $J_{AW}$ and predicts the appearance of a second faster timescale mode at small $J_{AW}$, which is seen in simulations (black curve in B). 
   (D) The power spectral density of activity fluctuations in the adapting model for the same coupling pairs as (B). The inset shows the oscillatory periods corresponding to the peaks in the spectral density.
   (E) The power spectral density for the adapting model at $J_{AW}=0.35$ and different adaptation rates $k_R$, with $k_R/k_B$ fixed. Slower adaptation leads to slower oscillations (inset). To compare our simulated dynamics to experimental timescales we set the characteristic spin flipping time to $\omega_0^{-1}=10$~ms, following Ref.~\cite{keegstra2026spontaneousswitching}.
   }
   \label{fig:fluctuations}
\end{figure*}

\subsection{Allosteric couplings influence activity fluctuations and correlations   }

We next explore how the strengths of the allosteric couplings within the array impact temporal fluctuations and correlations in signaling activity. The coupling $J_{AA}$ in our model is reasonably well constrained by the Pi\~nas et al. data~\cite{pinas2022hexamericrings} (when $f=0$, $J_{AA}$ is the only coupling parameter available to fit cooperativity and sensitivity measurements), but $J_{AW}$ and $J_{WW}$ are not uniquely determined. Specifically, we can obtain the same Hill coefficient along an entire curve in the $(J_{AW}, J_{WW})$-space, shown in Fig.~\ref{fig:fluctuations}A (here $H=27$ at $f=1$). While the sets of couplings along this line have nearly identical static response properties, their dynamics differ dramatically. 

Focusing on the fully filled lattice ($f=1$) for simplicity, we simulated activity time series and computed the power spectral density $S(\nu) = |\tilde a(\nu)|^2$, where $\tilde a(\nu)$ is the Fourier transform of the activity trajectory. Spectral densities for the non-adapting model are shown in Fig.~\ref{fig:fluctuations}B for various pairs of couplings from Fig.~\ref{fig:fluctuations}A. Extracting the correlation time $\tau_\mathrm{cor}$ from the power spectra (see Methods), we find a non-monotonic dependence of $\tau_\mathrm{cor}$ on $J_{AW}$ (Fig.~\ref{fig:fluctuations}C, \emph{upper}). Despite these lattices having identical cooperativity (Hill coefficient), their correlation times can vary by nearly two orders of magnitude. Furthermore, extreme couplings [i.e. $(J_{AW}, J_{WW}) = (0.15,1.7)$, black curve] can also lead to multiple competing correlation times in the dynamics. 

Our mean-field theory predicts (see Methods) the diverging dominant correlation time and the appearance of a second short-timescale correlation in the limit $J_{AW}\to0$ (Fig.~\ref{fig:fluctuations}C, \emph{lower}). Here the long timescale emerges because the system is composed of homogeneous $a$- and $w$-lattices, with only weak coupling between them. The mean-field theory, however, breaks down when local $a$-$w$ correlations dominate ($J_{AW}\gg J_{AA}, J_{WW}$) and hence does not capture the increase in correlation time in this regime; improving the theory to account for local correlations will be the topic of future work.

In the model for adapting cells, there are additional dynamics and timescales introduced by the adaptation dynamics (methylation and demethylation). As demonstrated previously, this adaptive feedback combined with sufficient cooperativity can lead to noisy oscillations in the activity of the array~\cite{mello2004effectsreceptor}. We find that (at fixed cooperativity) the coupling strengths $(J_{AW},J_{WW})$ have considerable impact on the power spectra and activity oscillation period $T_\mathrm{osc}$ in the adapting model (Fig.~\ref{fig:fluctuations}D). Similar to correlation times in the non-adapting model, extreme couplings (large and small $J_{AW}$) can also considerably increase the oscillation period $T_\mathrm{osc}$ and can  introduce additional relaxation timescales into the dynamics (small $J_{AW}$). %
Finally, for fixed coupling, a decrease (increase) in the adaptation rate scales the oscillation period up (down), (Fig.~\ref{fig:fluctuations}E). The oscillatory timescale $T_\mathrm{osc}$ is thus determined by a combination of the net array cooperativity, the internal allosteric coupling strengths, and the adaptation rate.

\begin{figure*}
   \includegraphics[width=0.98\textwidth]{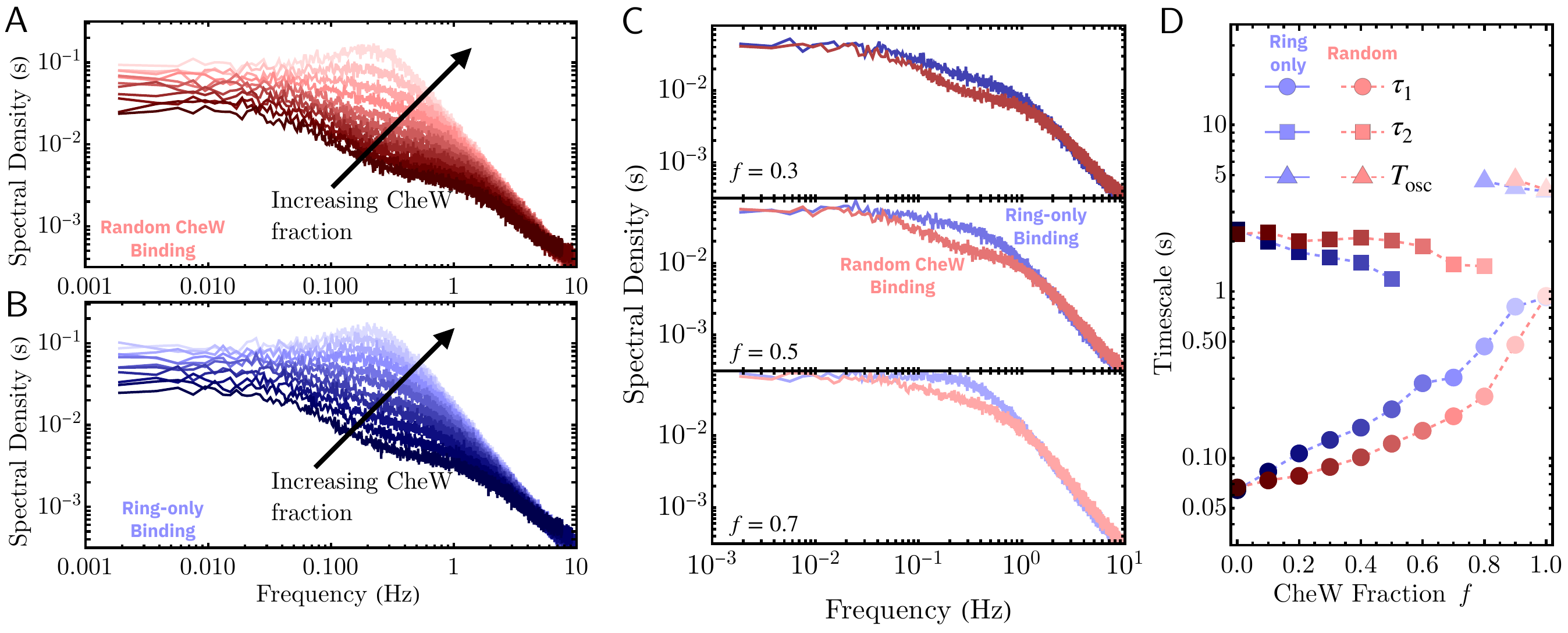}\\
   \caption{\emph{CheW participation modulates dynamic modes of the structured lattice.} 
   (A-B) The power spectral density of activity fluctuations in the adapting model as a function of CheW fraction $f$ (dark: $f=0$ to light: $f=1$) for randomly distributed CheW (A) and ring-only CheW lattices (B). 
   (C) Comparison of the power spectra for $f=0.3,0.5,0.7$ between the two spatial distributions of CheW.
   (D) Correlation and oscillatory timescales as a function of CheW fraction $f$. We find regions with multiple correlation times (small $f$), a single correlation time (intermediate $f$ in ring-only arrays), and oscillatory correlations (large $f$). In all panels we use: $(J_{AW}, J_{WW})=(0.35,0.375)$, spin flipping timescale $\omega_0^{-1}=10$~ms, $k_R =0.0015 \omega_0$, and $k_B/k_R$ tuned so that $A_0=1/2$.
   }
   \label{fig:fluctuationsCheWlevel}
\end{figure*}

\subsection{CheW level and assembly tune the array between multiple dynamic regimes}

In the analysis above, we fixed $f=1$ to study the impact of internal array couplings on dynamics. Since CheW expression is experimentally tunable and changes the effective coupling in the lattice (Eqs.~\ref{eq:Hill_nonadapt} and \ref{eq:K12}, Fig.~\ref{fig:RingvsRandom}), we next ask how the CheW level influences activity dynamics and fluctuations at fixed coupling. We simulated activity time series over a range of CheW fractions $f$, for both random and ring-only CheW placement. We focus on the adapting lattice model (with $A_0=1/2$), since the adaptive feedback gives rise to a richer range of dynamical properties.

Figs.~\ref{fig:fluctuationsCheWlevel}A and B show the power spectra for random and ring-only lattices respectively, for $f=0,0.1,\dots, 1$ (lighter curves correspond to higher $f$). Fig.~\ref{fig:fluctuationsCheWlevel}C compares the two spatial distributions of CheW directly for $f=0.3$, 0.5, and 0.7. As CheW fraction decreases, the system transitions from the adaptation-induced oscillatory fluctuations seen in Fig.~\ref{fig:fluctuations}D to relaxational fluctuations with one or two characteristic timescales. Across the full range of $f$, we found the activity correlations were captured well by one of three forms: $A e^{-|t|/\tau_1}$, $A e^{-|t|/\tau_1}+B e^{-|t|/\tau_2}$, or $A e^{-|t|/\tau_1}\cos(2\pi t/T_\mathrm{osc})+B e^{-|t|/\tau_2}$ (see Methods for the corresponding power spectra). The timescales extracted from power spectra fits are shown in Fig.~\ref{fig:fluctuationsCheWlevel}D. 

The spatial distribution of CheW placement on the array determines how the lattice transitions between these different dynamic regimes as the scaffolding protein is added. The random CheW lattices exhibited two resolvable correlation times for all $f<0.9$, with oscillations emerging when $f\geq0.9$. On the other hand, the ring-only lattices exhibited just a single resolvable mode at intermediate $0.6\leq f \leq 0.7$ and oscillations appeared for $f \geq 0.8$, below the threshold for random placement, due to the enhanced cooperativity of complete hexameric rings. While both CheW arrangements exhibited two timescales, the ring-only timescales lay between the random ones: $\tau_2^\mathrm{random}>\tau_2^\mathrm{ring}>\tau_1^\mathrm{ring}>\tau_1^\mathrm{random}$ (Fig.~\ref{fig:fluctuationsCheWlevel}D).

These results can be understood in terms of classical adaptive cooperative sensing. At low CheW fraction $f$, activity fluctuations exhibit fast (lattice) and slow (adaptation) relaxation modes. Increasing cooperativity (gain) makes activity more sensitive to adaptation, slowing the collective relaxation of the lattice and increasing the effective adaptation rate. This eventually causes the two relaxation rates to converge, leading to an oscillatory mode~\cite{mello2004effectsreceptor} at large $f$. The effective coupling grows faster with $f$ in ring-only lattices (Fig.~\ref{fig:RingvsRandom}) since complete hexamers provide the maximum local cooperativity, giving rise to oscillations at lower $f$ and accelerating the convergence of the relaxation rates so that they cannot be resolved even before the oscillatory transition. Our model illustrates the rich temporal dynamics exhibited by multi-component structured lattices. The emergence of slow oscillations with CheW expression (at a level dependent on the assembly of the array) provides a clear prediction for future FRET measurements of kinase activity fluctuations in single cells.

\section{Discussion}

\subsection{Probing array assembly dynamics and allostery via CheW expression and activity fluctuations}

Our study provides a quantitative modeling framework for understanding how an extended multi-component signaling complex, the chemosensory array, can tune the cooperativity and sensitivity of its response by varying the abundance of an internal scaffolding component, CheW, as observed in recent experiments~\cite{pinas2022hexamericrings}. Beyond capturing this existing data, our model makes testable predictions for how array responses and fluctuations depend on the microscopic details of CheW hexamer assembly and coupling, motivating future experiments that measure signaling behavior across a range of CheW expression levels.

In particular, we found that response properties depend considerably on the statistics of CheW placement in the array (Fig.~\ref{fig:RingvsRandom}). Measurements of response at additional sub-native CheW expression levels would therefore help reveal the nature of CheW hexamer assembly dynamics; the slope of the Hill coefficient or $K_{1/2}$ at low expression is particularly sensitive to the degree of randomness underlying ring formation. While available time-averaged data could not uniquely determine the coupling strengths in our model, we found that activity dynamics depend strongly on these internal couplings (Fig.~\ref{fig:fluctuations}). CheW expression and the array assembly pathway further modulate the dynamics between different relaxation and oscillatory modes (Fig.~\ref{fig:fluctuationsCheWlevel}). Measuring fluctuations in kinase activity via single-cell FRET, including across cells with different CheW expression levels, would therefore provide strong constraints on the allosteric coupling strengths between array components and the CheW spatial distribution. Observing the predicted set of dynamical transitions as a function of CheW expression would provide further support for our structure-based approach to modeling the chemosensory array.

\subsection{CheW-mediated structural diversity and tunable sensitivity}

It is also worth considering our results in the context of the broader diversity of chemosensory array structures observed across bacterial species~\cite{muok2020atypicalchemoreceptor} and even within \emph{E. coli}~\cite{burt2021alternativearchitecture}. For example, \emph{E. coli} minicells and \emph{Treponema denticola} arrays adopt p2 symmetry, containing only a single CheA-CheW hexamer type, in contrast to the mixture of CheA-CheW and CheW-only hexamers present in the p6 arrays studied here. Differently structured arrays have been suggested to arise from differences in membrane curvature and in CheA:CheW stoichiometry across species~\cite{muok2020atypicalchemoreceptor,burt2021alternativearchitecture}. Our results point to a further possibility: that these different array architectures encode functional trade-offs. 
For example, p2 arrays, with their more consistent and tightly coupled structure, may attain higher sensitivity and cooperativity, at the cost of being less adaptable to changes in protein expression or environmental conditions. By contrast, the p6 arrays studied here remain functional across a wide range of CheW expression levels and can tune their sensitivity and cooperativity simply by varying CheW hexamer participation.

This tunability may be functionally relevant even within a single species. Substantial phenotypic variability in chemotactic signaling and behavior has been observed among genetically identical bacteria, and populations can undergo non-genetic adaptation during growth and migration~\cite{moore2021nongeneticdiversity,moore2024signalintegration, vo2025nongeneticadaptation}. Our results suggest that stochasticity in CheW expression and array assembly, along with that of other array components, offers a plausible mechanism for generating this phenotypic variability. Linking heterogeneity in array structure to heterogeneity in chemotactic response is therefore a promising direction for future modeling work that could help identify the microscopic mechanisms underlying these population-level phenomena.

\subsection{Long-time structural adaptation}

While the presence of CheW rings in principle allows for adaptivity of the array, it remains an open question whether bacteria actually exploit this mechanism. Experiments by Frank and Vaknin~\cite{frank2013prolongedstimuli} and Piñas et al.~\cite{pinas2016sourcehigh}, measuring kinase activity and receptor conformations via anisotropy of homo-FRET signals, have shown that prolonged stimuli slowly reorganize the array, loosening receptor trimer packing and gradually decreasing kinase activity. Piñas et al.~\cite{pinas2016sourcehigh} attribute this change, at least partly, to loosening of the interface-2 interaction between core units, though disassembly of CheW hexamers likely also contributes to the observed structural changes. Modeling the coupled dynamics of CheW hexamer assembly/disassembly and array signaling is therefore an exciting direction for future theoretical work. Our model predicts that if CheW participation in the array changed slowly with sustained stimulus, this would considerably alter kinase dynamics in addition to causing the observed drift in average activity~\cite{frank2013prolongedstimuli, pinas2016sourcehigh}.

\subsection{Scaffold-tuned cooperativity beyond CheW}

Our results point toward a general principle for the study of scaffold-like components that lack catalytic or ligand-binding activity of their own. Since their contribution to signaling cannot be isolated through biochemical or single-molecule measurements of the component itself, structure-based modeling provides the most accessible path toward revealing their functional role. CheW exemplifies this paradigm: its activity cannot be measured directly, and its function only becomes apparent through its effect on the collective response of the array. 

Analogous scaffold-tuned cooperativity appears in other large signaling systems~\cite{good2011scaffoldproteins}. Among bacteria, \emph{Vibrio cholerae} has chemosensory arrays that contain two scaffolding proteins, CheW and CheV, whereas the catalytic protein CheA is very sparsely incorporated \cite{yang2018baseplatevariability}; the functional implications of this scaffold-dominated structure are unknown.  In cardiac and skeletal muscle, ryanodine receptor (RyR) channels self-assemble into extended lattices independently of any accessory protein~\cite{hu2005depletionfkbp}, but the accessory protein FKBP12 tunes the packing and cooperative gating of channels already in the lattice, inducing coupled gating in which neighboring channels open and close in near-unison~\cite{marx1998coupledgating,marx2001coupledgating}. The yeast mating pathway is another example: this system does not form an extended lattice, but the scaffold protein Ste5 organizes components of the MAPK signaling cascade via tethering. While Ste5 has no catalytic activity of its own, its self-dimerization and ligand-triggered conformational switching substantially change the ultrasensitivity of the downstream MAPK response~\cite{bhattacharyya2006ste5scaffold,takahashi2008membranelocalization,good2011scaffoldproteins}.

In each of these systems, as in the chemosensory array, a small non-catalytic component sets the effective coupling, cooperativity, or activity bias of the primary signaling machinery. A structure-based, quantitative modeling approach like ours could help reveal how that component's abundance, localization, or conformational state tunes the collective response.

\section{Outlook}

The structure-based model introduced here provides a first step toward understanding how the detailed protein interaction structure of the chemosensory array shapes its function. By explicitly modeling the activity states of both core units and CheW within hexameric rings, we can quantitatively explain how response cooperativity and sensitivity change as CheW expression is varied---a result that could not be obtained from a coarser, non-structural model. More broadly, this work suggests that CheW may serve as a functional tuning knob that adjusts array sensitivity and dynamics via long-timescale structural adaptation and that could contribute to phenotypic variability in chemotactic sensitivity across bacterial populations.

The model nonetheless simplifies the real array considerably, most notably by coarse graining the internal structure of the core signaling unit. Developing models that resolve this internal structure would help explain how interactions between different receptor types influence signaling in mixed core units and arrays \cite{sourjik2004functionalinteractions, mello2005allostericmodel,li2014selectiveallosteric}, and how response properties change as receptor stoichiometry and spatial arrangement evolve over the cell cycle \cite{frank2026singlearraymeasurements}. Incorporating the phosphorylation cycles underlying chemosensory signal propagation into a nonequilibrium structure-based model \cite{hathcock2023nonequilibriumallosteric,hathcock2024timereversalsymmetry} would further enable direct study of the interplay between nonequilibrium driving and array structure. Together, these directions will provide a quantitative understanding of how the array's structure governs its signaling function.

Beyond the chemosensory array, we expect this modeling strategy to generalize to other large signaling systems in which accessory scaffold components contribute to the cooperativity and activity of catalytic or receptor subunits. Outside of bacterial chemotaxis, RyR/FKBP12 clusters during calcium release \cite{marx1998coupledgating,marx2001coupledgating} and the Ste5 scaffold in the yeast mating MAPK pathway \cite{bhattacharyya2006ste5scaffold,takahashi2008membranelocalization} are suitable targets for our approach. In each case, the contribution of scaffolding proteins to signaling is difficult to measure in isolation. Hence, structure-based modeling offers a necessary complement to structural and single-molecule experiments for isolating the scaffold's functional role.

\section{Acknowledgements}
We thank Qiwei Yu for helpful discussions. Y.Z. would like to thank Prof. Junhua Yuan and Prof. Rongjing Zhang for their discussions and support during the course of this work. D.H. thanks the CCB at the Flatiron Institute for support while (a portion of) this research was carried out. D.H. is supported by Natural Sciences and Engineering Research Council of Canada (NSERC) Discovery Grant RGPIN-2026-07389. The Flatiron Institute is a division of the Simons Foundation.

\bibliography{ref_chemotaxis_clean}

\clearpage

\section{Methods}

\subsection{Lattice Simulations}

\subsubsection{Monte Carlo Simulations}

To represent the chemosensory array structure, we embed the baseplate Kagome structure of the array in a square lattice: lattice points $(i,j)$ with $i,j$ both even represent the holes in the Kagome lattice, where $w$-spins participating in the CheW hexamers are located (if $f>0$), while all other lattice points are occupied by a core unit $a$-spin. The $a$-spins are coupled on the square edges and additionally spins at $(i,j)$ where one of $i$ or $j$ is even are coupled to the $a$-spin at $(i+1,j-1)$. The $w$-spins are coupled to their neighbors (if any) in the hexamer and to the corresponding $a$-spin on the lattice. To reduce boundary effects in the simulation, we use periodic boundary conditions; lattices of size $L=36$ ($N_a=972$, $N_w=1944f$) are used throughout our study.

At intermediate CheW levels $0<f<1$, we populate the CheW hexamers to model two scenarios for CheW distribution and array assembly: (1) full rings are randomly placed at open sites in the Kagome lattice or (2) CheW are placed randomly across all possible hexamer sites in the array. In each case the number of CheW units fills a fraction $f$ of all available hexameric sites.

For measurements of activity dose-response curves, the methylation level of each core unit is fixed to be constant since methylation and demethylation rates are slow compared to kinase response rates. In non-adapting QEQE cells methylation is fixed at $m=2$ for all core units. In adapting cells, to determine the distribution of methylation across core units we simulate methylation and demethylation dynamics (see following section) to measure the steady-state distribution. We use the exact spatial distribution of methylation levels (i.e. we do not resample from the measured $P(m)$), which is then frozen for sampling the average activity.

Once the CheW positions and methylation levels are determined we simulate the activity using a standard Metropolis Monte Carlo algorithm~\cite{newman1999monte}, with spin flips accepted according to changes in energy given by the Hamiltonian, Eq.~(\ref{eq:model_hamiltonian}). Spins are sequentially updated in random order and data is recorded every ``lattice time unit'' $T_L = N_a+N_w$ flips. 

To measure dose-response curves, for each ligand concentration we discard an initial transient of at least  $3000\,T_L$ and record activity data for at least $7000\,T_L$. Dose-response curves are averaged over at least 100 (200) realizations for non-adapting (adapting) cells. Response parameters $H$ and $K_{1/2}$ are determined by fitting a Hill curve $1-[L]^H/([L]^H + K_{1/2}^H)$ to the normalized activity response. While this definition of the Hill coefficient allows for the closest comparison to experiments, we note that for $H\gg1$, the Ising response is not well fit by the Hill curve. This underestimates $H$ compared to the derivative definition $-2 d\log\langle a \rangle/d\log[L]$, but does not impact our qualitative results.

For measuring power spectra in Figs.~\ref{fig:fluctuations} and \ref{fig:fluctuationsCheWlevel}, we record activity data for at least $500{,}000\,T_L$ for each parameter set. Trajectories are split into segments of length $50{,}000\,T_L$ and the power spectra are averaged (over 100 realizations) to reduce noise. Non-adapting correlation times (Fig.~\ref{fig:fluctuations}B) are fit to the power spectrum for an exponential autocorrelation, $S_1(\nu) =A [1+(2 \pi \nu \tau_1)^2]^{-1}$ or double exponential $S_2(\nu) =A [1+(2 \pi \nu \tau_1)^2]^{-1}+B [1+(2 \pi \nu \tau_2)^2]^{-1}$ (for $J_{AW}=0.15$) with $\tau_1$ reported as $\tau_\mathrm{cor}$ in Fig.~\ref{fig:fluctuations}C.  Oscillation periods in the model of adapting cells (Fig.~\ref{fig:fluctuations}D--E) are fit to a power spectrum with oscillatory and exponential components, $S_3(\nu) =A [1+(\nu-T_\mathrm{osc}^{-1})^2 (2 \pi \tau_1)^2]^{-1} + A[1+(\nu+T_\mathrm{osc}^{-1})^2 (2 \pi \tau_1)^2]^{-1} + B [1+(2 \pi \nu \tau_2)^2]^{-1}$. For Fig.~\ref{fig:fluctuationsCheWlevel}, we fit each of $S_1$, $S_2$, and $S_3$ to the power spectrum and select the correlation model with the largest adjusted $R^2$. If $R^2$ values of two models differ by less than 0.1\%, we select the model with fewer parameters. To convert lattice dynamics to experimental timescales we assume a fundamental spin flipping timescale $\omega_0^{-1}=10$~ms~\cite{keegstra2026spontaneousswitching}.

\subsubsection{Adaptation dynamics}

We simulate adaptation following an approach similar to Refs.~\cite{mello2004effectsreceptor, keegstra2026spontaneousswitching}. Core units in the active state $a=1$ and $m>0$ are demethylated with rate $k_B$ while units in the inactive state $a=0$ with $m<4$ are methylated with rate $k_R$. After each spin flipping trial, the core unit is also methylated (demethylated)  with probability $k_R \omega_0^{-1}$ ($k_B \omega_0^{-1}$). We set $k_R=0.0015\omega_0 = 0.15\,\mathrm{s}^{-1}$~\cite{keegstra2026spontaneousswitching} and choose $k_B$ so that the average activity is $A_0=1/2$. Due to boundary effects from the discrete methylation state space $m\in \{0,\dots, 4\}$, $k_B/k_R$ is generally greater than 1 to achieve the half-activity adapted state. Finally, in Fig.~\ref{fig:fluctuations}E we studied the impact of slower and faster adaptation rates $k_R=0.08\,\mathrm{s}^{-1}$ and $0.3\,\mathrm{s}^{-1}$ while keeping $k_R/k_B$ fixed.

\subsection{Mean-Field Theory}

To formulate a mean-field theory for our structure-based Ising model, we start by determining the average coordination numbers in the structured lattice: $z_{aa}=4$, $z_{aw}=2f$, $z_{wa}=1$, and $z_{ww}=2$ (ring-preferred assembly) or $2f$ (random assembly), where $z_{xy}$ is the average number of $y$-spins coupled to a given $x$-spin. The mean fields are thus,
\begin{equation}
    \begin{split}
        h_s &= -z_{aa} J_{AA} \langle s \rangle - z_{aw} J_{AW} \langle \sigma \rangle + \frac{1}{2} h_\mathrm{tot}\\
        h_\sigma &= -z_{ww}J_{WW} \langle \sigma \rangle - z_{wa} J_{AW} \langle s \rangle,
    \end{split}
\end{equation}
where $h_\mathrm{tot} = h_L([L])+h_M(m)$ is the total ligand and methylation field applied to the core unit. In the following calculations we use the ring-preferred case $z_{ww}=2$; the same results hold for random assembly with the substitution $J_{WW} \to f J_{WW}$.

\subsubsection{Perturbation theory -- Non-adapting cells}
For mutants with fixed methylation level across all receptors, the mean-field formulation leads to the following self-consistency relations,
\begin{equation}\label{selfConsistency}
\begin{split}
    \langle s \rangle &= \tanh \bigg ( 4 J_{AA} \langle s \rangle + 2 f J_{AW} \langle \sigma \rangle -\frac{1}{2} h_\mathrm{tot} \bigg )\\
    \langle \sigma \rangle &=\tanh \bigg (2 J_{WW} \langle \sigma \rangle + J_{AW} \langle s \rangle \bigg).
\end{split}
\end{equation}
For QEQE mutants, the methylation field $h_M(m)$ is large enough in magnitude that the cluster is active at zero ligand $-h_M(m)\gg 1$ yet small enough that at saturating ligand, the activity is fully suppressed $h_\mathrm{tot}([L]\to\infty) = \log(K_a/K_i) + h_M \gg 1$. Thus, the half-maximum of the response occurs at $\langle a \rangle = 1/2$ or $\langle s \rangle = \langle \sigma \rangle = h_\mathrm{tot}=0$. Considering a perturbation $h_\mathrm{tot}=\delta h$, $\langle \sigma \rangle =\delta \sigma $, $\langle s \rangle =\delta s $ around this point, to linear order the self-consistent equations become:
\begin{equation}
\begin{split}
    \delta s =& 4 J_{AA} \delta s+ 2 f J_{AW}\delta \sigma - \frac{1}{2}\delta h\\
    \delta \sigma =& 2 J_{WW} \delta \sigma + J_{AW} \delta s.
\end{split}
\end{equation}
Solving for $\delta s$, $\delta \sigma$, we find,
\begin{equation}
\begin{split}
    \delta s &= -\frac{1}{2}\frac{1-2  J_{WW}}{(1- 4 J_{AA})(1-2  J_{WW}) - 2 f J_{AW}^2 }\delta h\\
    \delta \sigma &=-\frac{1}{2}\frac{J_{AW}}{(1- 4 J_{AA})(1-2  J_{WW}) - 2 f J_{AW}^2 }\delta h.
\end{split}
\end{equation}
The coefficient of $\delta h$ in the solution for $\delta s$, combined with the factor $\chi_h$, from the sensitivity of the ligand field $h_L$ at $K_{1/2}$ produces the Hill coefficient given in Eq.~(\ref{eq:Hill_nonadapt}).

\subsubsection{Relaxation Dynamics -- Non-adapting cells}

We can compute the temporal correlations for our model by assuming Glauber-like dynamics for the mean-field theory (assuming zero external field),
\begin{equation}
\begin{split}
        \omega_0^{-1} \partial_t \langle s\rangle &= -\langle s \rangle + \tanh \bigg ( 4 J_{AA} \langle s \rangle + 2 f J_{AW} \langle \sigma \rangle \bigg )\\
    \omega_0^{-1} \partial_t \langle \sigma\rangle &=   -\langle \sigma \rangle +\tanh \bigg (2 J_{WW} \langle \sigma \rangle + J_{AW} \langle s \rangle \bigg).
\end{split}
\end{equation}
Computing the Jacobian eigenvalues at $s=\sigma=0$ we find correlation timescales
\begin{equation}
\begin{split}
    \tau_\mathrm{mft,\pm}^{-1} = \omega_0 \Big(1&-2J_{AA}- J_{WW} \\
    &\pm \sqrt{2 f J_{AW}^2 + (J_{WW}-2 J_{AA})^2}\Big).
\end{split}
\end{equation}
Fig.~\ref{fig:fluctuations}C shows these timescales as a function of $J_{AW}$ for $f=1$, $J_{AA}=0.1$ with $H$ given in Eq.~(\ref{eq:Hill_nonadapt}) fixed.

\subsubsection{Perturbation theory -- Adaptation-proficient cells}

For adapting cells, we have a quenched random field $h_\mathrm{tot}(m)$ with methylation distribution $P(m)$ across the core units. Thus we have a mean-field self-consistency relation conditioned on the value of methylation,
\begin{equation}
    \langle s|m \rangle = \tanh \bigg ( 4 J_{AA} \langle s \rangle + 2 f J_{AW} \langle \sigma \rangle -\frac{1}{2} h_\mathrm{tot}(m) \bigg ).
\end{equation}
Averaging over the distribution of methylation produces a self-consistency relation involving only $\langle s \rangle = \sum_m \langle s|m \rangle P(m)$ and $\langle \sigma \rangle$,
\begin{equation}\label{selfConsistencyRandomField}
\begin{split}
    \langle s \rangle &= \sum_m P(m) \tanh \bigg ( 4 J_{AA} \langle s \rangle + 2 f J_{AW} \langle \sigma \rangle -\frac{1}{2} h_\mathrm{tot}(m) \bigg )\\
    \langle \sigma \rangle &=\tanh \bigg (2 J_{WW} \langle \sigma \rangle + J_{AW} \langle s \rangle \bigg).
\end{split}
\end{equation}
Expanding near zero background $[L]=0$ or $h_L([L])=\delta h_L$, we have $\langle s \rangle = s_0 - \delta s$, $\langle \sigma \rangle = \sigma_0 - \delta \sigma$. We assume the zeroth-order solution satisfies the self-consistency relations. The linear perturbation then becomes,
\begin{equation}
\begin{split}
    \delta s &= \frac{1}{2}\chi_s (8 J_{AA} \delta s +4 f J_{AW} \delta \sigma + \delta h_L) \\
    \delta \sigma &= \chi_\sigma (2 J_{WW} \delta \sigma + J_{AW} \delta s),
\end{split}
\end{equation}
where
\begin{equation}
\begin{split}
    \chi_s &= \sum_m P(m) \sech^2 \left [4 J_{AA} s_0 + 2 f J_{AW} \sigma_0 -\frac{1}{2} h_M(m)\right ] \\
    \chi_\sigma &= \sech^2 (2 J_{WW} \sigma_0 + J_{AW} s_0),
\end{split}
\end{equation}
are respectively the susceptibilities of $s$ and $\sigma$ spins at $[L]=0$. Eliminating $\delta \sigma$ and solving for $\delta h_L$ leads to 
\begin{equation}\label{eq:dhLinearApprox}
    \delta h_L = 2 \delta s \left (\frac{1}{\chi_s} -4 J_{AA} - \frac{2 f J_{AW}^2 \chi_\sigma}{1-2 J_{WW} \chi_\sigma}\right).
\end{equation}
If the activity is $A_0$ at $[L]=0$, we are interested in the point at which the activity drops to $A_0/2$ or $\delta s=(s_0+1)/2=A_0$. Plugging this into Eq.~(\ref{eq:dhLinearApprox}) and using $h_L([L]) \approx [L]/K_i$ for $[L] \ll K_i \ll K_a$, we arrive at the result quoted in the main text, Eq.~(\ref{eq:K12}).

\subsubsection{Hill coefficient in adapting MWC model}

To provide intuition for the invariance of the Hill coefficient in the adaptation-proficient cells, we can use the MWC model of the chemosensory cluster~\cite{shimizu2010modulargradientsensing,tu2013quantitativemodeling}. For a cluster size of $N$ and energy difference $E_s$ between the active and inactive states (see Ref.~\cite{shimizu2010modulargradientsensing}) the average activity is, 
\begin{equation}
    \langle a \rangle = \left(1+ e^{E_s}\left(\frac{1+[L]/K_i}{1+[L]/K_a} \right)^N \right)^{-1}.
\end{equation}
In previous work~\cite{shimizu2010modulargradientsensing} adaptation is encoded in methylation dependence of $E_s(m)$, which is tuned so that the activity attains the adapted level at zero ligand concentration, $\langle a \rangle([L]=0)=A_0$. The concentration at which the activity reaches half-maximum $A_0/2$ is then
\begin{equation}
    K_{1/2} = \frac{K_i (\mathcal{A}_N-1)}{1-\mathcal{A}_N K_i/K_a}
\end{equation}
where $\mathcal{A}_N = [(2-A_0)/(1-A_0)]^{1/N}$. Computing the Hill coefficient at this point, $H = -2\eval{\partial \log \langle a \rangle / \partial \log [L]}_{[L]=K_{1/2}}$, leads to
\begin{equation}
    H = \frac{(2-A_0)N(\mathcal{A}_N-1)(1-\mathcal{A}_N K_i/K_a)}{(1-K_i/K_a)\mathcal{A}_N}
\end{equation}
which approaches $H\approx (2-A_0) \log[(2-A_0)/(1-A_0)]$ as quoted in the main text. Note that for biological values of $K_i$ and $K_a$ (Table~\ref{tab:parameters}), we find $N\gtrsim 5$ is sufficient for convergence to the limit above.

\vspace{1em}

\end{document}